\documentclass[]{aa}
\usepackage{amsmath,amssymb}
\usepackage{graphics,latexsym,epsfig,txfonts,graphicx,natbib}
\usepackage[hyperindex,breaklinks]{hyperref}

\bibpunct{(}{)}{;}{a}{}{,}

\begin{document}

\title{A bias in cosmic shear from galaxy selection:\\ results from ray-tracing simulations}

\author{J.\ Hartlap\inst{1}\thanks{E-mail: {\tt hartlap@astro.uni-bonn.de}} \and S. Hilbert\inst{1,2} \and P.\ Schneider\inst{1} \and H.\ Hildebrandt\inst{3}}

\institute{Argelander-Institut f\"ur Astronomie, Universit\"at Bonn, Auf dem H\"ugel 71, D-53121 Bonn, Germany\and
  Max Planck Institute for Astrophysics, Karl-Schwarzschild-Str. 1, 85741 Garching, Germany\and
  Leiden Observatory, Leiden University, Niels Bohrweg 2, NL-2333 CA Leiden, The Netherlands}

\date{Received  / Accepted }
\authorrunning{Hartlap et al.}
\titlerunning{A bias in weak lensing from galaxy selection}

\abstract{}
{We identify and study a previously unknown systematic effect on cosmic shear measurements, caused by the selection of galaxies used for shape measurement, in particular the rejection of close (blended) galaxy pairs.}
{We use ray-tracing simulations based on the Millennium Simulation and a semi-analytical model of galaxy formation to create realistic galaxy catalogues. From these, we quantify the bias in the shear correlation functions by comparing measurements made from galaxy catalogues with and without removal of close pairs. A likelihood analysis is used to quantify the resulting shift in estimates of cosmological parameters. }
{The filtering of objects with close neighbours (a) changes the redshift distribution of the galaxies used for correlation function measurements, and (b) correlates the number density of sources in the background with the density field in the foreground. This leads to a scale-dependent bias of the correlation function of several percent, translating into biases of cosmological parameters of similar amplitude. This makes this new systematic effect potentially harmful for upcoming and planned cosmic shear surveys. As a remedy, we propose and test a weighting scheme that can significantly reduce the bias. }
{}
{}

\maketitle
%---------------------------------------------------------------------------------------

\section{Introduction}
In preparation for upcoming and planned large cosmic shear surveys, such as PAN-STARRS \citep{panstarrs}, KIDS\footnote{\tt http://www.eso.org/sci/observing/policies/\\PublicSurveys/sciencePublicSurveys.html} or Euclid \citep{euclid}, it is vital to find and quantify possible sources of systematic effects that can hamper the full exploitation of the information contained in these large data sets. A number of such effects have already been identified. The most fundamental problem on the observational side is to obtain unbiased estimates of the shapes of galaxies. The difficulty of this has been demonstrated, for example, by the STEP programme \citep[][]{STEP1,STEP2} and the GREAT08 challenge \citep[][]{GREAT08}, where several shape measurement methods have been tested on mock data. Further, it is crucial to obtain reliable photometric redshifts to obtain an accurate redshift distribution of the galaxy sample under consideration, which is needed for accurate theoretical predictions, and also to allow the construction of redshift bins for shear tomography \citep{hearin2010,bernstein2010,Ma2006}.
Intrinsic alignments of physically close galaxies and shape-shear alignments probably constitute the most severe physical contaminant of the cosmic shear signal. The first can be reduced by removing physically close pairs \citep{kingII,kingIIb,heymansII,takadaII}, whereas the influence of the latter can either be removed by the so-called nulling technique \citep{joachimiNulling,joachimiNulling_photoz} or by self-calibration \citep{zhangGI,JoachimiBridle09}.
Another physical contamination is caused by the magnification effect. Density fluctuations in the foreground can, depending on the slope of the galaxy number count, either enhance or deplete the number of background galaxies, thus correlating the density field in the foreground with the galaxy distribution that is used to estimate its shear field \citep{schmidt_lensingbias}.

In order to achieve percent-level accuracy, the difference between shear and reduced shear also needs to be taken into account. Theoretical predictions for the shear correlation functions can be obtained from the matter power spectrum with relative ease \citep[e.g.][]{bs}, but for the computation of the actually observable reduced shear correlation functions it is necessary to include higher-order corrections to the shear power spectrum \citep{white05,krause09}.

Finally, the process of parameter estimation requires great care as well. For example, the likelihood of the shear correlation functions has been shown to be significantly  non-Gaussian \citep{hartlapICA,schneiderCCF}. This may also apply to other two-point statistics derived from the correlation functions. Furthermore, even if a Gaussian likelihood is assumed, the cosmology dependence of the covariance matrix of the statistics under consideration should be taken into account, as was shown in  \citet{eifler08}. Neglecting these issues could introduce non-negligible biases to estimates of cosmological parameters.

In this paper, we add to this list a systematic effect that leads to a biased estimate of the shear correlation functions.
This bias is due to the fact that the ellipticity of a galaxy cannot be estimated reliably when its light distribution overlaps with that of a close neighbour. Therefore, it is common to discard  pairs of galaxies that appear too close together on the sky.  We argue that -- while allowing for clean estimates of galaxy shapes -- this practice has two adverse side effects: it changes the redshift distribution of source galaxies, and it correlates the lensing mass distribution in the foreground with the source galaxy distribution in the background.
Because of the latter issue, the product of the complex ellipticities of a randomly selected pair of galaxies no longer yields an unbiased estimate of the shear correlation function.

The article is organized as follows: after briefly reviewing the cosmic shear two-point statistics relevant for this paper in Sect.~\ref{sec:2ptstat}, we describe in Sect.~\ref{sec:rt} the ray-tracing simulations and semi-analytic galaxy formation models we use to create our mock galaxy catalogues.
In Sect.~\ref{sec:biasquant}, we quantify the bias in the shear correlation function using our simulation results for various choices of galaxy selection criteria. We then propose   a weighting scheme that can help to reduce the bias (Sect.~\ref{sec:weight}) and discuss the impact on cosmological parameter estimation (Sect.~\ref{sec:cospar}). We conclude the paper in Sect.~\ref{sec:conclusions}.

\section{The shear correlation functions} \label{sec:2ptstat}
Several statistics have been developed to capture the two-point information that is contained in the ellipticities of distant galaxies, such as the shear correlation functions
\citep[e.g.][]{kaiser92,crittenden}, the shear dispersion in circular apertures \citep[e.g.][]{kaiser92} or the aperture mass dispersion \citep[e.g.][]{schneider_map97,schneidermap}. Recently, \citet{cosebi} have proposed the so-called COSEBIs, which allow for a clean E-/B-mode decomposition given the shear correlation functions on a finite interval. These statistics are all related to the power spectrum of the weak lensing convergence \citep[see, e.g.,][]{crittenden,SvWM}.
Regarding actual measurements, the shear correlation functions $\xi_\pm$ are the most convenient of these statistics, since they can be estimated with relative ease from real data sets, even in the presence of gaps and masked regions. Any other two-point statistic of interest is therefore usually computed from an estimate of the shear correlation functions.
 
A practical estimator for the shear correlation functions is given by \citep[e.g.][]{schneidermap}
\begin{equation} \label{eq:correst}
  \hat\xi_\pm (\theta) = \frac{\sum_{i\neq j} \rho_i\,\rho_j\,\left(\epsilon_{{\rm t},i;j}\epsilon_{{\rm t},j;i} \pm \epsilon_{\times,i;j}\epsilon_{\times,j;i}\right)\,\Delta_{ij}(\theta)}{\sum_{i\neq j} \rho_i\,\rho_j \,\Delta_{ij}(\theta)}\;,
\end{equation}
where the sum runs over all pairs of galaxies, located at the angular positions $\vec\theta_{i}$.  The complex ellipticity of the $i^{\rm th}$ galaxy is denoted by $\epsilon_{i}$, and its tangential and cross components with respect to the line joining it to the $j^{\rm th}$ galaxy are given by $\epsilon_{{\rm t},i;j}$ and $\epsilon_{\times,i;j}$, respectively. The symbol $\Delta_{ij}(\theta)$ is equal to one if the angular separation $\theta$ of the  $i^{\rm th}$ and $j^{\rm th}$ galaxies lies in the bin centred on $\theta$, and vanishes otherwise. Finally, the $\rho_i$ are weights assigned to the galaxies. For the purpose of this work, it is convenient to write them as $\rho_i = m_i\,s_i$, where $s_i$ is a statistical weight that, for example, reflects the quality of the shape estimate. The ``selection weight'' $m_i$ is zero if the galaxy is too close to its nearest neighbour to allow for a reliable measurement of its shape, and unity otherwise.

\section{Ray-tracing simulations}\label{sec:rt}
Our ray-tracing simulations are based on the dark matter distribution in the Millennium Simulation \citep[MS,][]{mrpaper}. The cosmological parameters used for the MS (${\Omega_{\rm m} = 0.25}$, ${\Omega_{\rm DE}=0.75}$, ${\Omega_{\rm b}=0.045}$, ${\sigma_8=0.9}$, $h=0.73$, ${w_0=-1.0}$, $ {n_{\rm s}=1.0}$) also define the fiducial cosmological model used throughout the paper.

We have used the ray-tracing code described in \citet{mrrt} to obtain $32$ realisations of a $4\times4\;{\rm deg}^2$ field, thus covering $512\;{\rm deg}^2$ in total.
The matter distribution along the backwards light cone of the observer is obtained by the periodic continuation of simulation snapshots of increasing redshift. It is then divided into slices of a thickness of $\approx 100\,h^{-1}\,{\rm Mpc}$, which are subsequently projected onto lens planes. The periodic repetition of structures along the line of sight (l.o.s.) is prevented by choosing a l.o.s. direction that is tilted with respect to the boundaries of the simulation box.
The advantage of this technique in comparison to the random transformation approach is that the matter distribution is continuous across slice boundaries and that large-scale correlations extending beyond the redshift slices are maintained. 
The code follows a set of light rays, which form a grid on the first lens plane (the image plane), through the array of lens planes. At the same time, the Jacobian matrices of the lens mapping from the observer to the lens planes are computed using a recursion formula.

\begin{figure} 
\resizebox{\hsize}{!}{\includegraphics[]{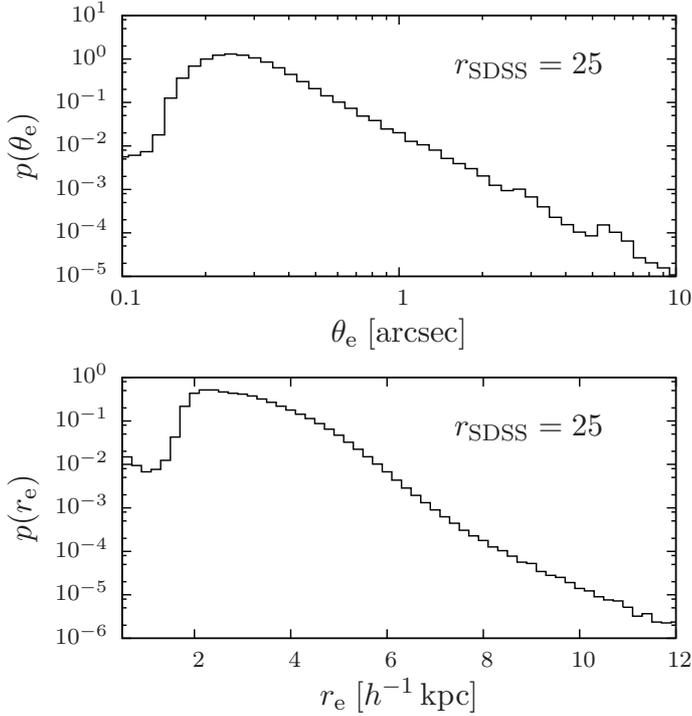}}
\caption{Distribution of galaxy radii in the simulated catalogue with $r_{\rm SDSS}=25$. \textit{Upper panel}: angular radius (no seeing), \textit{lower panel}: comoving physical radius.}
\label{fig:radii}
\end{figure}

To create realistic, lensed mock galaxy catalogues, we combine the ray-tracing with the semi-analytic model of galaxy formation by \citet{deluciaSAM}, making extensive use of the public Millennium Simulation database \citep{mrdb,lemson06}. We use the method outlined in \citet{mrrt} to obtain the lensed positions and observed magnitudes (taking the magnification due to lensing into account) for all galaxies in the semi-analytic model with $M_{\rm stellar}\geq10^9\,h^{-1}\,M_\odot$.
In addition, the galaxy formation model yields the masses of the disk and spheroidal (henceforth bulge) component and the disc radius $r_{\rm disc}$ (which can be zero). As described in more detail in \citet{hilbert08}, we complement this with an estimate of the comoving radius of the spheroidal component of the galaxy given by
\begin{equation}
	r_{\rm bulge} = 0.54\,(1+z)^{0.55}\left(\frac{M_{\rm bulge}}{10^{10}\,h^{-1}\,M_\odot}\right)^{0.56}\;h^{-1}\,{\rm kpc}\;,
\end{equation}
which combines the size distribution of galaxies measured by \citet{shen2003} and the redshift evolution of galaxy sizes found by \citet{trujillo2006}.
Each galaxy is then assigned an effective radius \mbox{$r_{\rm e} = \max(r_{\rm disc},r_{\rm bulge})$}. The angular diameter of the galaxy is given by $\theta_{\rm e} = \sqrt{\mu}\, r_{\rm e}/f_K(w)$, where $f_K(w)$ is the comoving angular diameter distance to the galaxy, and $\mu$ is the lensing magnification at the position of the galaxy.
The resulting distributions of angular and comoving galaxy radii for a simulated galaxy survey with a magnitude cut of $r_{\rm SDSS}=25$ are shown in Fig.~\ref{fig:radii}.

We construct our catalogues for cosmic shear measurements by selecting galaxies brighter than three different cuts in the SDSS $r$-band ($r_{\rm SDSS} = 24,\, 25,\, 26$). Unless otherwise stated, we assume that these cuts are the same as the limiting magnitude of the survey, $r_{\rm SDSS}^{\rm lim}$. For comparison, however, we will also consider the case where $r_{\rm SDSS}<r_{\rm SDSS}^{\rm lim}$. Since we assume that the check for overlapping light distributions is done before the galaxies are selected for shape measurement, galaxies that are brighter than $r_{\rm SDSS}$ and have faint close neighbours with magnitudes between $r_{\rm SDSS}$ and  $r_{\rm SDSS}^{\rm lim}$ are  removed from the lensing catalogue as well.

Furthermore, we use two criteria to identify pairs of objects whose projected angular separation $\theta$ is too small for obtaining reliable shape measurement of the individual galaxies:
\begin{itemize}
\item According to the first criterion, two galaxies at $\vec\theta_1$ and $\vec\theta_2$ with angular separation $\theta=|\vec\theta_1 - \vec\theta_2|$ are both removed from the catalogue if $\theta < \alpha(\theta_{{\rm e},1}'+\theta_{{\rm e},2}')$. Here, the effective angular radii are given by
 \begin{equation}
		\theta_{{\rm e},i}' = \sqrt{\mu \,\left[r_{\rm e}/f_K(w_i)\right]^2 + \theta_{\rm see}^2}\;,
  \end{equation}
	where $\theta_{\rm see}$ is the size of the seeing disk. The parameter $\alpha$ can be chosen arbitrarily to tune the strictness of the selection criterion and to compensate for inaccuracies of our modelling of the galaxy radii. Since this criterion depends on the half-light radius of the galaxies, we henceforth denote it with ``HLR''.
\item The second criterion (called ``FIX'' criterion) is similar to what is used for, e.g.,  the CFHTLS \citep[see also][]{vWME,maoli2001}. It uses a fixed angular separation threshold: if a pair of galaxies fulfils $\theta < \theta_{\rm fix}$, one of the two galaxies is selected at random and removed from the catalogue. The rationale for doing this is the following: even though the light distribution of the remaining galaxy is still affected by the light of the removed neighbour, the resulting error of the shape estimate should be uncorrelated with any other galaxy that remains in the catalogue and should just add to the noise. In addition to this, we remove all galaxies that are members of obviously severely blended pairs by applying the HLR criterion with $\alpha=1$. We find, however, that the effect of this second step on the properties of the resulting galaxy catalogue is generally sub-dominant.
\end{itemize}

The choice of the selection criterion and its parameters most likely depends on the quality of the data at hand and the shape measurement pipeline used. In general, one wishes to retain as many galaxies as possible while keeping the bias caused by isophote overlap below a certain threshold.

We remark that a multitude of variants of these selection criteria can be conceived, where for example for the FIX criterion, not a random galaxy is removed from a close pair, but the galaxy with the lowest signal-to-noise ratio (SNR); a related possibility would be to always keep close pairs when the SNR of one galaxy is considerably larger than the SNR of the second. Using these variants is expected to lead to minor quantitative, but not to qualitative changes of the results presented in the following sections.

In Tables \ref{tab:ngal_hlr} and \ref{tab:ngal_fix}, we list the galaxy number densities after applying the HLR and FIX criteria, respectively, for various values of $\alpha$,  $\theta_{\rm see}$ and $\theta_{\rm fix}$. The values given in parentheses are the fractional decrease of the number density compared to the unfiltered galaxy catalogue. As expected, the deeper the survey and the more restrictive the criterion, the more galaxies are removed, since the probability of overlap is proportional to the projected galaxy density and the square of the threshold radius of the selection criterion.

\begin{table}
	\caption{Galaxy number densities for the HLR criterion}
	\centering
	\begin{tabular}{c c l l l}
		\hline\hline
		$\alpha$ &  $\theta_{\rm see}$ & $\bar n_{\rm gal}(r_{\rm SDSS}=24)$ & $\bar n_{\rm gal}(r_{\rm SDSS}=25)$ & $\bar n_{\rm gal}(r_{\rm SDSS}=26)$ \\
		& $[\arcsec]$ & $[{\rm arcmin}^{-2}]$ & $[{\rm arcmin}^{-2}]$ & $[{\rm arcmin}^{-2}]$\\
		\hline
		 $1$ & $0.00$ & $11.3$ $(0.1\%)$ & $24.9$ $(1.2\%)$ & $49.9$ $(1.5\%)$\\
		 $1$ & $0.65$ & $11.0$ $(3.5\%)$ & $23.8$ $(5.6\%)$ & $46.2$ $(8.9\%)$\\
		 $2$ & $0.00$ & $11.0$ $(3.5\%)$ & $24.2$ $(4.0\%)$ & $48.2$ $(5.0\%)$\\
		 $2$ & $0.65$ & $10.2$ $(10.5\%)$ & $20.9$ $(17.1\%)$ & $36.7$ $(27.6\%)$\\
		 $3$ & $0.00$ & $10.6$ $(7.0\%)$ & $23.2$ $(7.9\%)$ & $45.7$ $(9.9\%)$\\
		 $3$ & $0.65$ & $10.2$ $(10.5\%)$ & $17.2$ $(31.7\%)$ & $26.8$ $(47.1\%)$\\
		\hline
	\end{tabular}
  \tablefoot{Values in parentheses are the fractional decrease of the galaxy number density compared to the values without removal of close pairs. These are $11.4$, $25.2$ and $50.7/{\rm arcmin}^2$ for $r_{\rm SDSS}=24,\,25\,$ and $26$, respectively.}
  \label{tab:ngal_hlr}
\end{table}

\begin{table}
	\caption{Galaxy number densities for the FIX criterion}
	\label{tab:galdens}
	\centering
	\begin{tabular}{c c l l l}
		\hline\hline
		$\theta_{\rm fix}$ &  $\theta_{\rm see}$ & $\bar n_{\rm gal}(r_{\rm SDSS}=24)$ & $\bar n_{\rm gal}(r_{\rm SDSS}=25)$ & $\bar n_{\rm gal}(r_{\rm SDSS}=26)$ \\
		$[\arcsec]$ & $[\arcsec]$ & $[{\rm arcmin}^{-2}]$ & $[{\rm arcmin}^{-2}]$ & $[{\rm arcmin}^{-2}]$\\
		\hline
		 $2.0$ & $0.65$ & $10.9$ $(4.4\%)$ & $23.3$ $(7.5\%)$ & $44.2$ $(12.8\%)$ \\
		 $3.7$ & $0.65$ & $10.3$ $(9.6\%)$ & $21.0$ $(16.7\%)$ & $36.7$ $(27.6\%)$\\
		 $5.0$ & $0.65$ & $9.8$ $(14.0\%)$ & $18.9$ $(25.0\%)$ & $30.5$ $(40.0\%)$\\
		\hline
	\end{tabular}
	 \tablefoot{Values in parentheses are the fractional decrease of the galaxy number density compared to the values without removal of close pairs. These are $11.4$, $25.2$ and $50.7/{\rm arcmin}^2$ for $r_{\rm SDSS}=24,\,25\,$ and $26$, respectively.}
  \label{tab:ngal_fix}
\end{table}

We compute the shear correlation functions from our simulated galaxy catalogues using Eq.~\eqref{eq:correst}. We obtain the observed galaxy ellipticities $\epsilon$ using the relation \citep{schneiderseitz95}
\begin{equation} \label{eq:ellpties} \epsilon=\begin{cases}
	\frac{\epsilon^{(\rm{s})} +g}{1+g^*\epsilon^{(\rm{s})}} & {\rm if}\; |g|\leq 1\\
	\frac{1+g\epsilon^{*(\rm{s})}}{\epsilon^{*(\rm{s})} +g^*} & {\rm if}\; |g|> 1\; ,
	\end{cases}
\end{equation}
where $g$ is the reduced shear obtained from the ray-tracing simulations and $\epsilon^{(\rm{s})}$ is the intrinsic ellipticity. For measuring the bias caused by the selection criteria described above, we set $\epsilon^{(\rm{s})}=0$ in Eq.~\eqref{eq:ellpties}. We include intrinsic ellipticities only when computing the covariance matrix of the shear correlation functions for the discussion in Sect.~\ref{sec:cospar}.

\section{The effect of object selection} \label{sec:biasquant}

\subsection{Effect on the redshift distribution}\label{sec:biasquant_1}

\begin{figure}
\resizebox{\hsize}{!}{\includegraphics[]{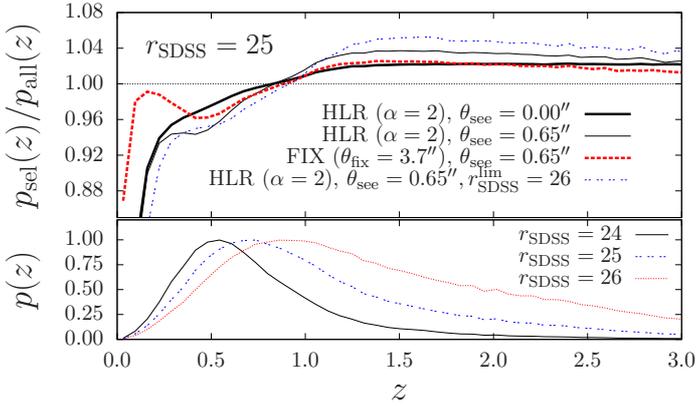}}
\caption{\emph{Upper panel:} Ratio of the redshift distributions for $r_{\rm SDSS}=25$ after and before object selection for the HLR criterion with $\alpha=2$ (thick solid line: $\theta_{\rm see}=0\farcs 0$, thin solid line: $\theta_{\rm see}=0\farcs 65$) and the FIX criterion with $\theta_{\rm fix}=3\farcs 7$ (thick dashed line). The case with a deeper limiting magnitude ($r_{{\rm SDSS}}^{\rm lim}=26$) than the magnitude cut for the lensing catalogue is represented by the double-dotted blue line. \emph{Lower panel:} Redshift distributions of the unfiltered galaxy catalogues with limiting magnitudes $r_{\rm SDSS}=24$ (solid line), $r_{\rm SDSS}=25$ (dashed line) and $r_{\rm SDSS}=26$ (dotted line).  }
 \label{fig:zratio}
\end{figure}

Using a selection criterion like the ones described in the previous section has two undesirable side effects: first, the redshift distribution of the filtered galaxy catalogue is different from the one before object selection. This is illustrated in Fig.~\ref{fig:zratio}, where we show in the lower panel the redshift distributions of our three mock surveys with different magnitude cuts without applying any selection criterion. The upper panel displays the ratios of the redshift distributions after and before object selection for the survey with $r_{\rm SDSS}=25$. We also consider the case of a limiting magnitude of the survey that is deeper than the cut used to define the sample of galaxies used for shape measurements ($r_{{\rm SDSS}}^{\rm lim}=26$, whereas $r_{{\rm SDSS}}=25$). Seeing only has very little effect on the results obtained with the FIX criterion, because the size of the seeing disk is typically much smaller than the fixed threshold radius $\theta_{\rm fix}$; we therefore only consider the case with $\theta_{\rm see} = 0\farcs 65$.

In all cases, the largest fraction of the galaxies is removed at low redshifts. These galaxies have the largest apparent radii and thus have the largest probability of isophote overlap. The amplitude of the deviations is highest for $r_{{\rm SDSS}}^{\rm lim} > r_{{\rm SDSS}}$, because additional pairs are removed in which one galaxy is from the magnitude range $[r_{\rm SDSS},r_{{\rm SDSS}}^{\rm lim}]$.
For the FIX criterion and the HLR criterion with seeing, a secondary dip occurs, approximately at the redshift of the peak of the redshift distribution.
This behaviour is due to the presence of angular clustering.

We illustrate this by constructing a simple analytical model, for which we subdivide the galaxies into redshift slices of width $\mathrm{d} z$. We assume that there are no angular cross-correlations between slices at different redshifts. Furthermore, we use a power-law model for the angular correlation function, so that it is given by $\omega(\theta; z, z' ) = A(z)\,\theta^{-\gamma}$ if $z=z'$ and $\omega(\theta; z, z' ) = 0$ otherwise. The galaxy radii are given in a deterministic fashion by $\theta_{\rm e}(z) = \sqrt{r_{\rm e}^2(z)/f_K[w(z)] + \theta_{\rm see}^2}$, where $r_{\rm e}(z)$ can be considered to be the mean radius of all galaxies in a thin redshift bin centred on $z$. The probability of finding a galaxy with redshift $z'$ within an annulus of radius $\theta$ and width $\mathrm{d} \theta$ around a galaxy at redshift $z$ is thus $\mathrm{d} p(z,z') = 2\pi\theta\,\mathrm{d}\theta\,[1+\omega(\theta; z, z' )] N(z')/\Omega$, where $\Omega$ is the total area of the survey, and $N(z')$ is the number of galaxies in the redshift slice centred on $z'$. Two galaxies have overlapping isophotes if they are closer than $\theta_{\rm eff}(z,z') =\theta_{\rm e}(z) +  \theta_{\rm e}(z')$ (corresponding to the HLR criterion with $\alpha=1$, which we choose here for simplicity). The total probability of overlap for two galaxies is given by the integral of $\mathrm{d} p(z,z')$ over a circle with radius $\theta_{\rm eff}(z,z')$. Finally, we obtain the total number of galaxies removed from the slice at $z$ by summing up the contributions from all redshift slices:
\begin{equation}
  \Delta N(z) = \frac{2\pi\,N(z)}{\Omega}\int \mathrm{d} z'\; N(z') \int_0^{\theta_{\rm eff}(z,z')}\mathrm{d}\theta\,\theta\,\left[1+\omega(\theta; z, z' )\right]\;.
\end{equation}
Simplifying and inserting our model for the correlation function, we obtain
\begin{equation} \label{eq:zdistratio}
  \frac{\Delta N(z)}{N(z)} = \frac{\pi}{\Omega}\left[ \int\mathrm{d} z'\,\theta_{\rm eff}^2(z,z')\,N(z') + \frac{A(z)\,N(z)}{2-\gamma} \theta_{\rm eff}^{2-\gamma}(z,z) \right]\;,
\end{equation}
where the second term accounts for the effect of galaxy clustering. We can simplify this even more by using a constant clustering amplitude and a redshift-independent radius $\vartheta$ for all galaxies, so that $\theta_{\rm eff}(z,z') = 2\vartheta$. We then find that
\begin{equation}
  \frac{\Delta N(z)}{N(z)} = \frac{\pi}{\Omega}\left[ 4\vartheta^2\,N_{\rm tot} + \frac{AN(z)}{2-\gamma}\,(2\vartheta)^{2-\gamma} \right]\;.
\end{equation}
We see that in the absence of angular correlations, a constant fraction of objects is removed from the total galaxy population, given only by the fraction of the total area covered by galaxies. This leaves the shape of the redshift distribution unchanged. Galaxy clustering increases the probability of overlapping isophotes. However, this is effective only for galaxies in the same redshift slice. The fraction of blended objects in a slice is proportional to $N(z)$ (and not $N_{\rm tot}$ as without clustering). This causes the secondary minimum seen in the upper panel of Fig.~\ref{fig:zratio} for those selection criteria where $\theta_{\rm e}(z)$ approaches a finite, constant value as $z$ increases. This is the case if seeing is present, as well as for the FIX criterion. If, on the other hand, $\theta_{\rm e}(z)$ is allowed to fall to zero, as for the HLR criterion without seeing, this suppresses the clustering term in Eq.~\eqref{eq:zdistratio}, because the galaxy radii (and thus $\theta_{\rm eff}^{2-\gamma}$) are already close to zero when $N(z)$ approaches its maximum.

The change of the redshift distribution is relevant if the redshifts of the individual galaxies are not available. In such cases, $p(z)$ is usually inferred from a sub-sample or a similar survey with either spectroscopic or photometric redshifts. In general, the objects used for these calibration samples are selected in a different way than the galaxies for the shear catalogue, and therefore the redshift distribution obtained in this way does not account for the change of $p(z)$ due to object selection. For upcoming lensing surveys incorporating photometric redshifts, this should be less of a concern, since in this case the redshift distribution of the galaxies in the shear catalogue can at least in principle be estimated directly.

\subsection{Density-dependent galaxy selection}

The second, and probably more severe effect of using selection criteria such as HLR and FIX arises because the selection is density-dependent. Since the galaxy distribution is correlated with the underlying density field, a mass overdensity in the foreground also implies an overdensity of galaxies. This in turn implies a higher probability of isophote overlap and thus of the removal of galaxies. Therefore, the fraction of galaxy pairs that can be formed from galaxies located behind overdensities is decreased relative to all galaxy pairs that contribute to the shear correlation function estimator for a certain angular separation bin. High-density regions are therefore effectively down-weighted compared to the case without object selection. The opposite is true for underdense regions: the probability that a galaxy behind the underdensity is filtered is reduced, and more pairs than for a region of average density contribute to the shear correlation functions. This re-weighting is further modified by the fact that the fraction of galaxies that is removed from the fore- and background of a specific lens is not constant (see Fig.~\ref{fig:zratio}). The ratio of the number of foreground-foreground and foreground-background pairs, which do not carry information about the lens, to the number of background-background pairs, where both galaxies have been sheared by the lens, depends on the lens redshift. If relatively more pairs in the background than in the foreground are removed, the signal of the lens is further suppressed (and vice versa).

The net effect of all this is that the shear correlation estimator given by Eq.~\eqref{eq:correst} is no longer unbiased. The reason for this is that the weights $\rho_i$ are no longer uncorrelated with the galaxy ellipticities, because the selection weights $m_i$ depend on the projected galaxy density through the mechanism described above.

\subsection{Simulation results}
We define the bias due to object selection as
\begin{equation} \label{eq:beta_zd}
  \beta_\pm^{z+\delta}(\theta) = \xi_\pm^{\mathrm{sel}}(\theta) - \xi_\pm^{\mathrm{all}}(\theta)\;,
\end{equation}
where $\xi_\pm^{\mathrm{sel}}$ are the shear correlation functions after filtering for close pairs, and $\xi_\pm^{\mathrm{all}}$ are the correlation functions computed from all galaxies in the field of view. The superscript ``$z+\delta$'' indicates that  $\beta_\pm^{z+\delta}$ includes the bias both due to the change of the redshift distribution and the density-dependent selection of galaxies.

\begin{figure}
\resizebox{\hsize}{!}{\includegraphics{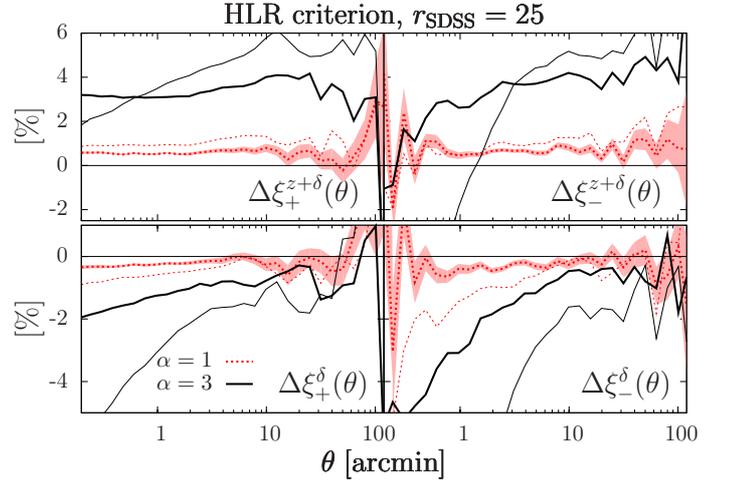}}
\caption{Fractional bias of the shear correlation functions for the HLR criterion, without (upper panels) and with (lower panels) correction for the change of the redshift distribution. Thick dashed lines are for $\alpha=1$, thick solid lines for $\alpha=3$ without seeing. For the respective thin lines a seeing of $0\farcs 65$ was assumed. The shaded region shows the $1\sigma$-error. For better visibility, it is shown only for the case of $\alpha=3$, $\theta_{\rm see}=0''$. The error bars for the other cases are very similar.  }
\label{fig:cfadd_hlr}
\end{figure}

\begin{figure}
\resizebox{\hsize}{!}{\includegraphics{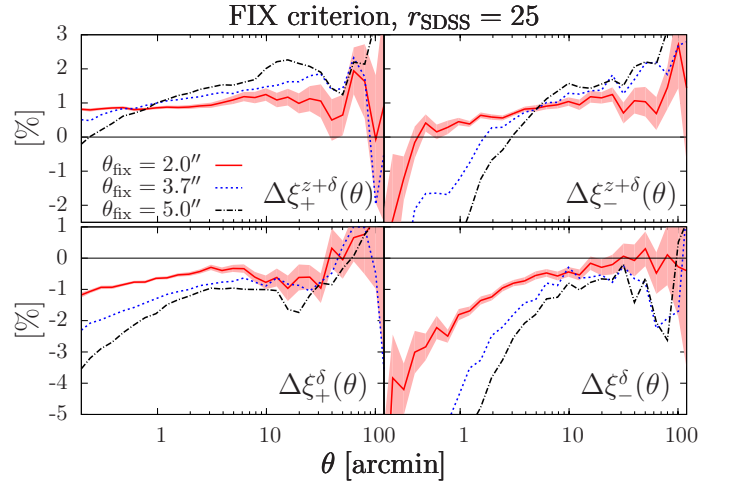}}
\caption{Same as Fig.~\ref{fig:cfadd_hlr}, but for the FIX criterion with $\theta_{\rm fix}=2''$ (solid red curves), $\theta_{\rm fix}=3.7''$ (short-dashed blue curves) and $\theta_{\rm fix}=5.0''$ (dot-dashed blue curves). }
\label{fig:cfadd_fix}
\end{figure}

In the upper panels of Figs. \ref{fig:cfadd_hlr} and \ref{fig:cfadd_fix}, we show the fractional bias
\begin{equation} \label{eq:deltaxi_zd}
 \Delta\xi_\pm^{z+\delta}(\theta) = \frac{\beta^{z+\delta}_\pm(\theta)}{\xi_\pm^{\mathrm{all}}(\theta)}\;
\end{equation}
for the HLR and FIX criterion, respectively. The error bars have been computed from the field-to-field variation between the $32$ ray-tracing realisations. For both criteria, $\xi_\pm^{\mathrm{sel}}$ is biased high by several percent on large scales, whereas on small scales this bias can become negative. The behaviour for large $\theta$ can be explained by the change of the redshift distribution, which gives more weight to high-redshift galaxies, which carry the strongest shear signal (see Fig.~\ref{fig:zratio}). On small scales, the negative bias begins to dominate due to the density-dependence of the way galaxy pairs are selected (see below).
As discussed before, seeing is only relevant for the HLR criterion and was therefore not considered in Fig.~\ref{fig:cfadd_fix}. 

\begin{figure}
\resizebox{\hsize}{!}{\includegraphics{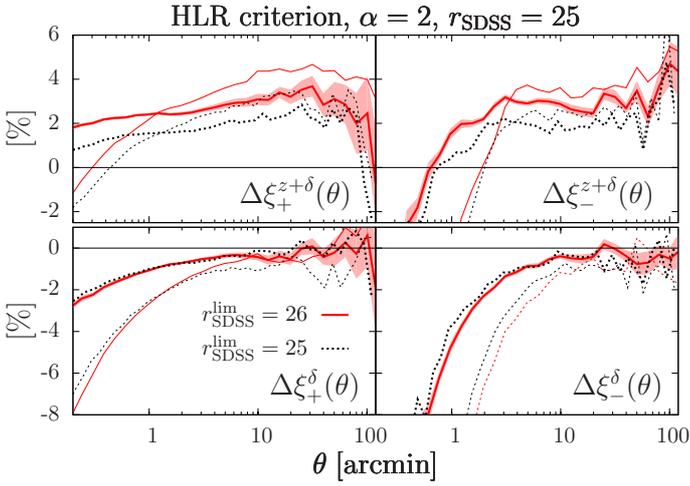}}
\caption{Comparison of the fractional bias of $\xi_\pm$ for a survey with limiting magnitude $r_{{\rm SDSS}}^{\rm lim}=26$ and a magnitude cut for the galaxies that are used for shape measurements of $r_{{\rm SDSS}}=25$ (solid lines), to the bias for a survey where $r_{{\rm SDSS}}^{\rm lim}=r_{{\rm SDSS}}=25$ (dashed lines). For both cases, the HLR criterion with $\alpha=2$ was used. Thick curves display the case without seeing, thin curves the case with  $\theta_{\rm see}=0\farcs 65$. }
\label{fig:cfadd_2cat}
\end{figure}

If photometric redshifts are available for all galaxies, the correct redshift distribution after object selection can be estimated, and one is only interested in the systematic effect caused by the density-dependence of the galaxy selection. To quantify this, we therefore would like to compare $\xi_\pm^{\mathrm{sel}}$ to fiducial correlation functions that were computed using the correct redshift distribution and with galaxy pairs selected in a fair way, i.e.~uncorrelated with the density field.
To this end, we take the unfiltered galaxy catalogues (the ones that led to $\xi_\pm^{\mathrm{all}}$), sort the galaxies into redshift bins, and randomly remove galaxies from each bin so that the resulting new galaxy catalogue has the same redshift distribution as the catalogue after applying one of the selection criteria. We denote the correlation functions computed from the new catalogues by $\xi_\pm^{\rm zcorr}$, and define the bias only due to the density-dependence of the galaxy selection process (indicated by the superscript ``$\delta$'') by
\begin{equation} \label{eq:beta_d}
	\beta_\pm^{\delta}(\theta) =\xi_\pm^{\rm sel}(\theta)-\xi_\pm^{\mathrm{zcorr}}(\theta)\;.
\end{equation}
Accordingly, the corresponding fractional bias is given by
\begin{equation} \label{eq:deltaxi_d}
	 \Delta\xi_\pm^{\delta}(\theta) = \frac{\beta_\pm^{\delta}(\theta)}{\xi_\pm^{\mathrm{zcorr}}(\theta)} \;.
\end{equation}

We show the simulation results for $\Delta\xi_\pm^{\delta}$ in the lower panels of Figs. \ref{fig:cfadd_hlr} and \ref{fig:cfadd_fix}. The bias is now consistent with being negative for all angular separations, as expected from the qualitative picture described in Sect.~\ref{sec:biasquant_1}. The effect is most severe for small $\theta$, whereas $\Delta\xi_\pm^{\delta}$ seems to asymptotically approach zero on large scales. Even after correcting for the change of the redshift distribution, the bias is of the order of several percent and therefore constitutes a potentially significant contaminant for present and future cosmic shear surveys.

\begin{figure}
\resizebox{\hsize}{!}{\includegraphics{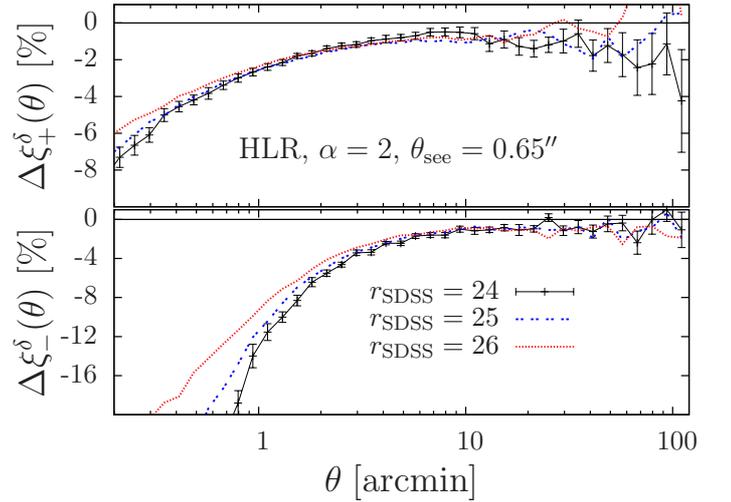}}
\caption{Fractional bias of the shear correlation functions for various survey depths using the HLR criterion with $\alpha=2$, $\theta_{\rm see}=0\farcs 65$, corrected for the change of the redshift distribution. Solid black lines with error bars: $r_{\rm SDSS}=24$; blue dashed line: $r_{\rm SDSS}=25$; red dotted line: $r_{\rm SDSS}=26$}
\label{fig:cfadd_magcut}
\end{figure}

We compare a survey with a limiting magnitude of \mbox{$r_{{\rm SDSS}}^{\rm lim}=26$} and a cut for the lensing catalogue of \mbox{$r_{{\rm SDSS}}=25$} to a survey with $r_{{\rm SDSS}}^{\rm lim}=r_{{\rm SDSS}}=25$ in Fig.~\ref{fig:cfadd_2cat}, using the HLR criterion with $\alpha=2$.
While $\Delta\xi_\pm^{z+\delta}$ increases by $\approx 1\%$ in the case with $r_{{\rm SDSS}}^{\rm lim}=26$ due to the change in redshift distribution (see also Fig.~\ref{fig:zratio}), no significant differences between the two surveys can be found if the correct redshift distribution is known. The reason for this is that although more galaxies are removed when the deeper limiting magnitude is used, the bias is primarily due to the change of the relative weights of over- and underdense regions in the correlation function estimator. These weights only depend on the relative change of the number of galaxy pairs behind such structures.
The same argument can be used to explain the results displayed in Fig.~\ref{fig:cfadd_magcut}, where we investigate the behaviour of the bias (taking the change of $p(z)$ into account) for various magnitude cuts (using the HLR criterion). We find that, for the magnitude cuts and the resulting redshift distributions of galaxies considered here, the fractional bias  $\Delta\xi_\pm^{\delta}$ depends only very little on the survey depth. The only notable difference occurs on small scales, where the bias for deeper surveys is slightly less severe than for shallower ones.

\section{Weighting scheme} \label{sec:weight}
The bias discussed in the previous sections is caused by the removal of galaxy pairs in a way that is correlated with the density field. This suggests that the bias could be reduced by increasing or decreasing the relative pair count behind over- and underdensities, respectively, to ``fair'' levels. Such a procedure would reduce both the bias due to the change of $p(z)$ and due to the density-dependent selection. We propose that, if photo-$z$ estimates are available also for the galaxies that have been filtered out, this can be achieved by identifying the nearest neighbour of a removed object on the sky, and doubling its weight for the correlation function computation (i.e. using it twice in the shear catalogue). We demand close proximity on the sky and in redshift to ensure that the shear of the neighbour is a reasonable proxy for the shear at the position of the filtered galaxy.

\begin{figure}
\resizebox{\hsize}{!}{\includegraphics{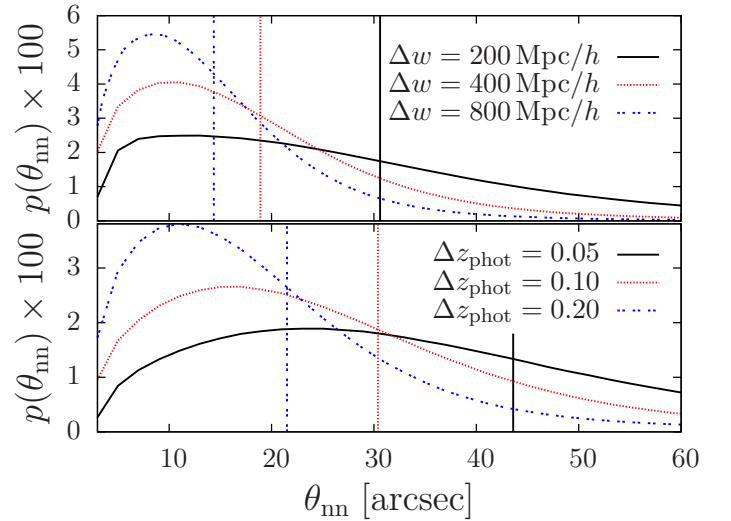}}
\caption{\emph{Upper panel:} distribution of the separation of rejected galaxies from their nearest accepted neighbour in a slice of thickness $\Delta w$; a magnitude cut of  $r_{\rm SDSS}=25$ and the HLR criterion with $\alpha=2$, $\theta_{\rm see}=0\farcs 65$ were used. \emph{Lower panel:} same as upper panel, but for slices with thickness given by $\Delta z_{\rm phot}$. Vertical lines indicate mean separations. }
\label{fig:nnhist}
\end{figure}

Since the geometrical lensing weight functions change relatively slowly with comoving distance, it is sufficient to choose an object from a redshift slice centred on the removed object with a certain width (a few hundred Mpc). This also helps finding a neighbour that is sufficiently close to the removed galaxy on the sky; clearly, the larger the slice width, the smaller is the projected nearest-neighbour distance of objects in the slice. This is illustrated in Fig.~\ref{fig:nnhist}, where we show the distribution of the distance from a rejected galaxy to its nearest neighbour for slices with widths specified in comoving distance (upper panel) or photometric redshift (lower panel).
The latter were obtained by simulating the photo-z accuracy in a
typical contemporary weak lensing survey. We use the recipe described
in \citet{HH2007,HH2009} to simulate a multi-colour
catalogue based on realistic distributions of redshift, spectral type, magnitude
and magnitude error, closely resembling the
CFHTLS-Wide. Finally, photo-$z$'s are estimated with the BPZ code \citep{benitez2000}. Comparisons of the simulated photo-$z$ accuracy to the one
obtained from real CFHTLS data \citep{erben2009} show good
agreement. For the thickest slices considered ($\Delta z_{\rm phot}=0.2$), the mean distance to the nearest neighbour is $44 \arcsec$, and decreases to $22 \arcsec$ for the slice with $\Delta z_{\rm phot}=0.05$. 

\begin{figure}
\resizebox{\hsize}{!}{\includegraphics{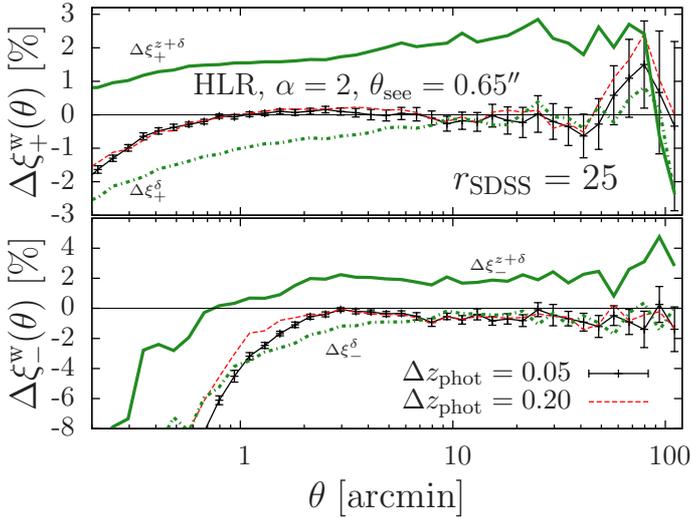}}
\caption{Fractional bias of $\xi_\pm$ using the weighting scheme described in Sect.~\ref{sec:weight}. Thin curves: fractional bias after applying the weighting scheme for slices with thickness $\Delta z_{\rm phot}$; thick solid green curve: $\Delta\xi_\pm^{z+\delta}$ without weighting scheme, thick dot-dashed green curve: $\Delta\xi_\pm^{\delta}$ without weighting scheme. Error bars were computed from field-to-field variation and for better visibility are shown only for one case.}
\label{fig:cfadd_weight}
\end{figure}

In Fig.~\ref{fig:cfadd_weight}, we compare the fractional bias of the correlation functions $\xi_\pm^{\rm w}$, measured after applying the weighting scheme,
\begin{equation} \label{eq:deltaxi_w}
 \Delta\xi_\pm^{\rm w} = \frac{\beta^{\rm w}_\pm(\theta)}{\xi_\pm^{\mathrm{all}}(\theta)}\; \;,
\end{equation}
where
\begin{equation} \label{eq:beta_w}
	\beta^{\rm w}_\pm(\theta)= \xi_\pm^{\rm w}(\theta)-\xi_\pm^{\mathrm{all}}(\theta)\;,
\end{equation}
to the original $\Delta\xi_\pm^{z+\delta}$  and $\Delta\xi_\pm^{\delta}$. We only consider slices defined in terms of photometric redshift; the corresponding results for slices with a given comoving thickness are very similar. The suggested procedure clearly reduces the bias, in particular on large scales.
Its performance degrades on scales below angular separations of $\approx 1\arcmin$. The reason for this is that the selection of the nearest neighbour effectively corresponds to a smoothing of the shear field with smoothing length comparable to the mean nearest-neighbour distance, because it is assumed that the shear of the removed galaxy is the same as the one of its substitute. Since $\xi_-$ is more sensitive to small-scale power than $\xi_+$, it should be particularly affected, as can indeed be seen in the lower panel of Fig.~\ref{fig:cfadd_weight}. The deviations seen for large $\theta$ are consistent with noise.
On the other hand, the actual width of the redshift slice and, related to this, the actual value of the mean nearest-neighbour distance have very little effect on the quality of the method, although there is a slight tendency for thicker slices to yield values of $\Delta\xi_\pm^{\rm w}$ that are larger by a few fractions of a percent. This means that the weighting scheme is relatively insensitive to the quality of the photometric redshifts. This is particularly important for those galaxies which have been filtered out because their light distribution is contaminated by a close neighbour, which also adversely affects the accuracy of their photo-$z$ estimates.

\section{Implications for cosmological parameters} \label{sec:cospar}

\begin{figure*}
\sidecaption
\includegraphics[width=12cm,angle=0]{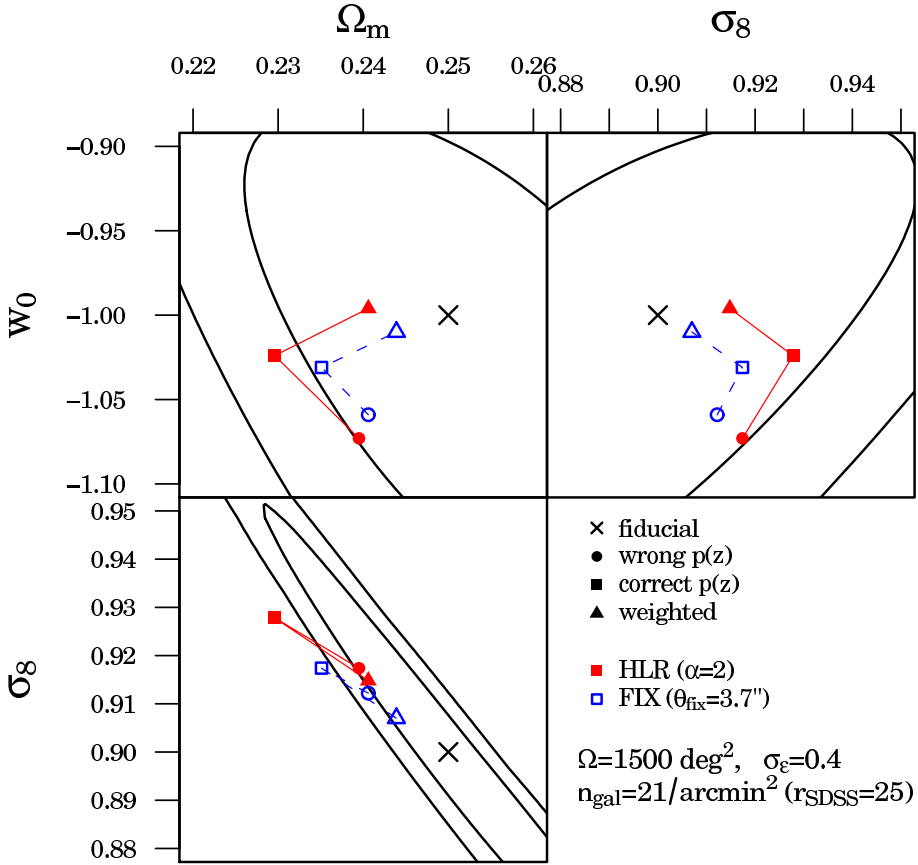}
\caption{Likelihood analysis of the bias caused by the removal of blended galaxies. Each panel shows the $1$- and $2\sigma$ confidence contours obtained by marginalizing over the remaining parameter (computed for the HLR criterion with $\alpha=2$ and $\theta_{\rm see}=0\farcs 65$). The fiducial parameter values are marked with crosses. Circles indicate the maximum likelihood estimates assuming that the true redshift distribution after object selection is unknown, squares show the maxima if the correct $p(z)$ is used, and triangles give the estimates after applying the weighting scheme of Sect.~\ref{sec:weight}. For the HLR criterion ($\alpha=2$; $\theta_{\rm see}=0\farcs 65$), filled symbols have been used, open symbols for the FIX criterion with $\theta_{\rm fix}=3.7''$. For better visibility, the estimates for each criterion type have been connected with a line.  }
\label{fig:params}
\end{figure*}

To illustrate the importance of the object selection bias for parameter estimation, we perform a likelihood analysis to fit for the parameters $\vec \pi = (\Omega_{\rm m},\sigma_8,w_0)$ (assuming a flat universe). Our fiducial cosmological model, denoted by $\vec\pi_0$, is that of the Millennium Simulation (see Sect.~\ref{sec:rt}). We use the galaxy sample with $r_{\rm SDSS}=25$ from our ray-tracing simulations for simulating a survey with an area of $1500\,{\rm deg}^2$. To each galaxy, we assign Gaussian ellipticity noise with dispersion $\sigma_{\epsilon}=0.4$. We consider correlation functions given on ten logarithmically spaced bins in the range from $1'$ to $80'$, and assume a Gaussian likelihood of the form
\begin{equation}
		p(\vec \xi | \vec \pi)  \propto \exp\left\{ -\frac{1}{2}\,[\vec \xi -\vec m(\vec \pi)]^{\rm t}\,\tens{C}^{-1}\,[\vec\xi -\vec m(\vec \pi)] \right\}\;.
\end{equation}
Here, $\vec \xi =\left[\xi_+(\theta_1),\,\ldots,\,\xi_+(\theta_n),\, \xi_-(\theta_1),\,\ldots,\,\xi_-(\theta_n)\right]^{\rm t}$ is the measured correlation function (see below), written in vectorial form, and $\vec m(\vec \pi)$ is the model prediction based on the three-dimensional matter power spectrum as given by \citet{smithps}. The covariance matrix $\tens C$  has been estimated from the field-to-field variation of the ray-tracing realisations. When computing its inverse, we correct for the bias caused by the noise in the estimate of $\tens{C}$ by using the correction factor described in \citet{hartlapcovar}.

Since we are only interested in the effect of removing close galaxy pairs and not in the bias caused by the mismatch of the theoretical model and the correlation functions in the Millennium Simulation \citep[see][]{mrrt}, we construct the data vectors $\vec \xi$ from the model for our fiducial set of parameters, $\vec m(\vec\pi_0)$, and the bias $\vec\beta$  due to object selection, measured from the ray-tracing simulations:
\begin{equation}
 \vec \xi = \vec m(\vec\pi_0) + \vec \beta\;.
\end{equation}
For $\vec\beta$, we consider three cases: (a) assuming no knowledge of the true redshift distribution after removing blended galaxies (i.e.~using $\beta_\pm^{z+\delta}$ from Eq.~\ref{eq:beta_zd}), (b) assuming that the change of the redshift distribution has been taken into account (using $\beta_\pm^{\delta}$ from Eq.~\ref{eq:beta_d}), and (c) assuming that the weighting scheme described in the previous section has been applied (using $\beta_\pm^{\rm w}$ from Eq.~\ref{eq:beta_w}).

 In Fig.~\ref{fig:params}, we show the results of this procedure for both the HLR criterion with $\alpha=2$ and $\theta_{\rm see}=0\farcs 65$, and the FIX criterion with $\theta_{\rm fix}=3\farcs 7$. In Tab.~\ref{tab:cparbias}, we show the fractional shifts of each cosmological parameter with respect to its fiducial value for the various cases.
As expected from the larger amplitude of the bias in the correlation functions for the HLR criterion (see Figs.~\ref{fig:cfadd_hlr} and \ref{fig:cfadd_fix}), the deviation of the maximum-likelihood points from the true values is generally larger for the HLR criterion than for the FIX criterion. In both cases, the parameter estimates are off by several percent. Interestingly, knowledge of the correct redshift distribution does not necessarily improve the parameter estimates, which is particularly striking in the $\Omega_{\rm m}$-$\sigma_8$-plane.
The bias of the maximum-likelihood estimates of all parameters is reduced by a significant amount if the proposed weighting scheme is applied, as could be expected from the likewise reduction of the bias in the shear correlation functions (see Fig.~\ref{fig:cfadd_weight}). While not being a perfect solution that would be accurate enough to be applied to planned large-area surveys, the method works sufficiently well to reduce the bias to a level that is well below the statistical errors for surveys like the CFHTLS (at least for the non-tomographic case considered here). Furthermore, a comparison of the correlation functions measured with and without the weighting scheme may be used to assess the importance of the object selection bias for a given survey.

%new results using xi_zcorr
\begin{table}[]
	\caption{Fractional bias of cosmological parameters}
	\centering
	\begin{tabular}{l l l l l l}
		\hline\hline
		 &  $\Omega_{\rm m}$ & $\sigma_8$ & $w_0$ \\
		\hline
		 \emph{HLR criterion}\\
		  $p(z)$ unknown & $4.2\%$&  $1.9\%$ & $7.3\%$\\
		 $p(z)$ known   & $8.2\%$&  $3.1\%$ & $2.4\%$\\
		 weighted		& $3.7\%$& $1.6\%$ & $-0.4\%$\\
		\hline
		 \emph{FIX criterion}\\
		 $p(z)$ unknown & $3.7\%$& $1.4\%$ & $5.9\%$\\
		 $p(z)$ known   & $6.0\%$& $2.0\%$ & $3.1\%$\\
		 weighted		& $2.4\%$& $0.8\%$ & $1.0\%$\\
		\hline
	\end{tabular}
   \label{tab:cparbias}
\end{table}

\section{Summary and conclusions} \label{sec:conclusions}
We have described a new, so far unconsidered systematic effect affecting the measurement of the shear correlation functions. The cause of the bias is the common practice of removing galaxies from the lensing catalogue that have very close neighbours, in order to avoid isophote overlap. While this filtering is necessary for obtaining clean shape estimates, it has two adverse effects on the correlation function estimate. The first consists in altering the redshift distribution of the galaxy catalogue. This is most important for low redshifts (where the angular sizes of galaxies are large) and (as a result of the angular clustering of the galaxies) near the peak of the redshift distribution. Second, such filtering predominantly removes galaxies that lie behind overdense regions. Therefore, fewer pairs of galaxies can be formed that carry the shear signal of the overdensity, effectively down-weighting it in the correlation function estimator. For similar reasons, underdense patches of the sky receive a higher weight.
As a result of this, the estimate of the shear correlation functions obtained from a galaxy catalogue, from which close pairs have been removed, is biased. 

In order to quantify this bias, we have run ray-tracing simulations through the Millennium Simulation in conjunction with a semi-analytic model of galaxy formation and observed scaling relations for the radii of the galaxies. We consider two different selection criteria, one that removes close pairs of galaxies closer than a certain threshold separation which depends on the radii of the galaxies, and the other removing one galaxy of a pair that is closer than a certain fixed threshold.
We find that the change of the redshift distribution due to filtering is of the order of several percent; however, this can in principle be dealt with if photometric redshifts are available for all galaxies.

The effect of the density-dependence of the galaxy selection varies with angular separation. We find that on scales of $\approx 1\arcmin$, the shear correlation functions are biased low by typically several percent; the bias decreases for larger angular separations. The bias seems to be almost independent of the survey depth. While seeing has essentially no effect on the bias when a fixed threshold radius is used to define close pairs, adding seeing to the simulations can significantly increase the bias for the selection criterion depending on the sizes of galaxies.

We note that the bias studied here is different from the effects of the clustering of source galaxies previously discussed in the literature \citep{bernardeauSC,SvWM}: the removal of close galaxy pairs creates an anti-correlation between foreground and background galaxies, and thus between the lensing matter distribution and the galaxies that are used to trace the shear field caused by the matter in the foreground. This induces clustering between galaxy populations that are widely separated in redshift, whereas the effect of \citet{SvWM} arises from the clustering of source galaxies that are at very similar redshifts and which need not be related to the dark matter distribution at all.

We have investigated the impact of the new systematic effect on estimates of $\Omega_{\rm m}$, $\sigma_8$ and $w_0$, assuming a flat universe and keeping all other parameters fixed. Irrespective of whether the correct redshift distribution is used or not, we find shifts of the maximum-likelihood estimators of several percent.
The situation can be significantly improved by using different weights for the galaxies that are eventually used for measuring the correlation functions. The weighting scheme consists of double-counting the nearest neighbour (from within a redshift slice with thickness of a few hundred $\mathrm{Mpc}$) of a galaxy that has been removed. This requires photometric redshift estimates to be available also for the galaxies that have been removed by the selection criterion. We find that the method works well even for slices as thick as $\Delta z_{\rm phot}=0.2$, so that the requirements on the quality of these redshift estimates are relatively low.
The weighting scheme restores the pair count to fair levels and substitutes the shear of the filtered galaxy with the shear of the nearest neighbour. The scheme is surprisingly independent of the actual width of the redshift slice and reduces the bias of the correlation function to levels of $\lesssim 1\%$  for angular scales ranging from $\approx 2\arcmin $ to $\approx 80\arcmin$. Accordingly, the bias of cosmological parameter estimates is also significantly reduced.

Given the amplitude of the bias of the shear correlation function, this new systematic effect  has the potential of being very significant. The weighting scheme we propose is a first step towards controlling it, but it probably lacks the accuracy necessary for the next generation of weak lensing experiments. 

\begin{acknowledgements}
We would like to thank Sherry Suyu, Tim Eifler, Benjamin Joachimi and Simon White for helpful discussions and input during the course of this project.
JH and SH acknowledge support by the Deutsche Forschungsgemeinschaft within the Priority Programme 1177 under the project SCHN 342/6 and the Transregional Collaborative Research Centre TRR 33 "The Dark Universe". HH was supported by the European DUEL RTN, project MRTN-CT-2006-036133. The Millennium Simulation databases used in this paper and the web application providing online access to them were constructed as part of the activities of the German Astrophysical Virtual Observatory.
\end{acknowledgements}

\bibliographystyle{aa}
\bibliography{references}

\appendix

% -------------------------------------------------------------------
\end{document}